\documentclass[11pt]{article}
\usepackage{amsfonts}
\usepackage[preprint]{acl}
\usepackage{listings}
\usepackage{amsmath}
\usepackage{times}
\usepackage{latexsym}
\usepackage{booktabs}
\usepackage[T1]{fontenc}

\usepackage[utf8]{inputenc}

\usepackage{microtype}

\usepackage{inconsolata}

\usepackage{graphicx}

\title{FlowEval: Reference-based Evaluation of Generated User Interfaces}

\author{
  Jason Wu \\
  Purdue University \\
  \texttt{jasonwu@purdue.edu}
  \And
  Priyan Vaithilingam \\
  Apple \\
  \texttt{priyan@apple.com}
  \And
  Eldon Schoop \\
  Apple \\
  \texttt{eldon@apple.com}
  \AND
  Jeffrey Nichols \\
  Apple \\
  \texttt{jwnichols@apple.com}
  \And
  Titus Barik \\
  Apple \\
  \texttt{tbarik@apple.com}
}

\begin{document}
\maketitle
\begin{abstract}
While large language models (LLMs) and coding agents are often applied to user interface (UI) development, developers find it difficult to reliably assess their proficiency in visual and interaction design.
Existing evaluations either rely on human experts, who can accurately assess usability by testing critical flows but are slow and costly, or on automated judges, which are scalable but less accurate and opaque.
We present FlowEval, a reference-based framework that measures whether a generated UI supports realistic interaction flows by comparing navigation traces from real websites to traces from generated analogs using reference-based similarity metrics (e.g., dynamic time warping).
In a small-scale study with expert UI evaluators, we show that reference-based metrics strongly correlate with human judgments, suggesting that they can provide scalable yet trustworthy evaluation for UI generation systems.
\end{abstract}

\section{Introduction}

Recent advances in large language models (LLMs) and coding agents have made it possible to automatically generate complex user interfaces (UIs), but it is difficult to automatically evaluate the quality of their outputs.
High-quality UIs must not only be visually appealing, but also functional and easy to use. 
Traditional approaches~\cite{nielsen1990heuristic,nielsen1994usability} to evaluating these factors rely on human expert review and testing, which can capture subtle design issues but are slow, costly, and difficult to scale.

One potential response is to use LLMs (or MLLMs) themselves as evaluators through prompting or fine-tuning on human responses, sometimes called ``LLM-as-a-judge''~\cite{zheng2023judging}.
Several such approaches have been specifically developed for assessing the quality of generated UI output~\cite{zhang2025artifactsbench,lin2025computeruse,li2025webdevjudge}.
However, such reference-free evaluators inherit the biases and failure modes of the underlying models and themselves require validation~\cite{li2025generation}. Because of this, it can be difficult for domain experts to shape or trust these scores as an evaluation signal.
\begin{figure}
    \centering
    \includegraphics[width=0.95\linewidth]{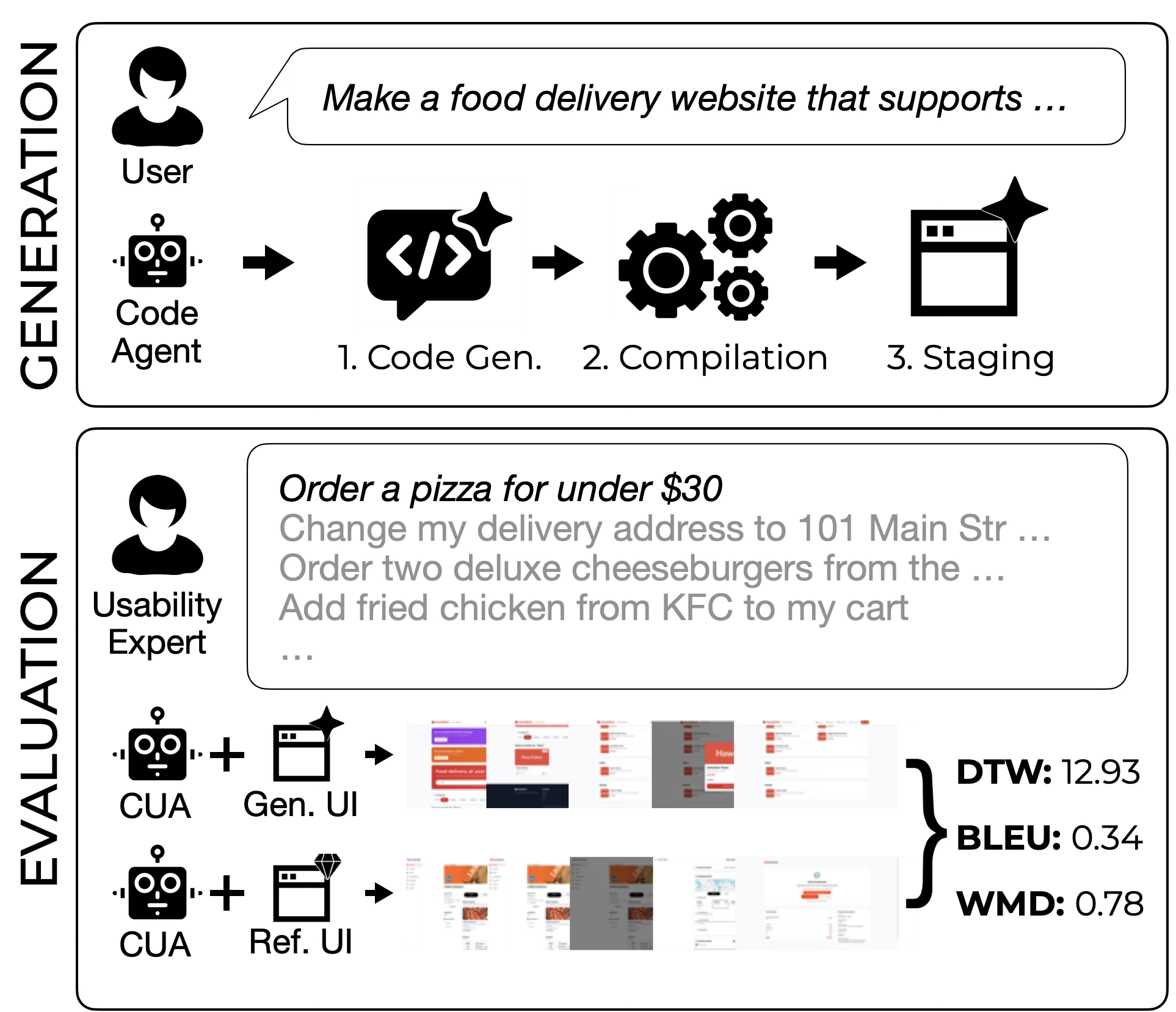}
    \caption{An overview of our evaluation approach. A coding agent generates an output, which is compared against a reference website. A computer use agent (CUA) produces and compares traces between generated and reference UIs.}
    \label{fig:herofigure}
\end{figure}
Reference-based approaches, which have long been used in NLP and other areas of machine learning, offer a promising alternative.
Compared to ``reference-free'' approaches, like automatic judges, they leverage ground-truth from known good examples, rather than relying solely on an evaluator model’s internal notion of what a good output should be, which helps disentangle failures of the evaluator from genuine problems in the evaluated content.
Domain experts can directly encode their subjective knowledge by curating reference designs and task sets for more targeted and bespoke evaluation, e.g., specific design criteria or adherence to a company's specific design language.
Finally, metrics computed with references are also more interpretable and enable more targeted debugging, which is helpful for complex design and interaction patterns.

We introduce FlowEval, a reference-based framework for evaluating interactive generated UIs.
FlowEval uses a set of predefined interaction traces, or ``flows,'' collected by performing tasks on a known high-quality UI and a generated analog to compute reference-based metrics.
We describe an automated implementation that generates interaction traces by running a computer use agent (CUA) using validated tasks sourced from navigation benchmarks.
We apply several types of reference-based metrics found in the NLP literature to compute similarity between traces collected from the reference and synthetic UIs.
Finally, we conduct an arena evaluation where human experts provided 429 ratings of LLM-generated websites and compared it to our automated assessment method.
Our results show that FlowEval's scores strongly correlate with human preferences and can provide more aligned rankings than a baseline MLLM judge approach used by previous work.

\section{Related Work}
Several benchmarks have been developed for assessing LLM code~\cite{Chen2021EvaluatingLL}, which often include front-end languages~\cite{peng-etal-2024-humaneval}.
Many focus on tasks with references such as functional code correctness and ``design-to-code'' tasks~\cite{si-etal-2025-design2code}.
However, it is difficult to automatically measure success for design quality.

Traditionally, this has relied on expert judgment~\cite{nielsen1994usability}, but this is difficult to scale.
Prior research has explored encoding expertise into manually crafted objective functions~\cite{todi2016sketchplore,oulasvirta2018aim}, symbolic models~\cite{ritter2000supporting,paterno2004concurtasktrees}, and guidelines~\cite{nielsen1990heuristic,shneiderman2016dtui} for humans and machines to follow.
Yet, it is hard to capture expert design knowledge in a form that is both comprehensive and broadly applicable.

Another approach is to engage a larger number of non-experts through crowdsourcing~\cite{Luther:2015:SAE:2675133.2675283} or model ``arenas''~\cite{designarena2025,vichare2025webdev}.
Because they cannot be run in a fully automated fashion, collected labels have, in turn, been used to improve automated approaches such as VLM judges for design aesthetics and ~\cite{duan2024uicrit,wu2024uiclip} and interaction flows~\cite{zhang2025artifactsbench,lin2025computeruse,li2025webdevjudge}.
However, they can be less accurate due to high rater disagreement~\cite{wu2024uiclip} and provide opaque scores.

Our work contributes to automated UI evaluation by retaining expert input through curated references, while using automated agents and metrics to enable scalable assessment.

\section{FlowEval}
We introduce FlowEval, a framework for evaluating UIs by generating interaction traces  from validated tasks and computing similarity to high-quality reference flows.
We describe our approach to i) generating interactive UIs that correspond to known references, ii) generating interaction traces using a CUA, and iii) applying three representative reference-based metrics to compute trace similarity.

\textbf{Source Data.} As a source of reference datasets and validated interaction tasks, we chose two CUA navigation benchmarks, although any expert-validated references and task flows can be used.
We chose i) WebVoyager~\cite{he-etal-2024-webvoyager}, which consists of 15 live websites and ii) REAL v2~\cite{garg2025real}, which provides 12 high-fidelity replicas for reproducible evaluation of stateful flows that are impractical to test on live sites (\textit{e.g.,} authenticated or transaction-dependent interactions).
WebVoyager contains a total of 643 tasks; but many of these tasks have a high degree of overlap, \textit{e.g.,} multiple tasks that book a flight for different destinations.
We distilled this into a smaller set by randomly sampling half of each website's tasks then manually removing repeated flows, resulting in 147 final tasks.
We also manually filtered REAL's 121 tasks, which results in 116 final tasks.

\subsection{UI Generation}
We evaluated five LLMs, which included state-of-the-art proprietary and open-source LLMs at the time of our experiments: GPT 5 Codex~\cite{openai2025gpt5_codex_addendum}, Gemini Pro 2.5~\cite{comanici2025gemini25}, Claude Sonnet 4.5~\cite{anthropic2025claude_sonnet_45}, Qwen3Coder 30B, Qwen3Coder 480B~\cite{qwen2025qwen3_coder}, GPT-OSS 20B, and GPT-OSS 120B~\cite{agarwal2025gpt_oss_model_card}.
We used each model's official coding agent, which improved performance over manual one-shot model prompting through multi-turn execution and tool-calling: OpenAI Codex, Gemini CLI, Claude Code,  and Qwen Code.

Each model was used to prompted to generate a UI in a ``one-shot'' manner using a React.js harness.
We modified a public prompt and harness used for UI benchmarking~\cite{designarena2025} to include a description of the website and a list of tasks that must be implemented (see Appendix).
Each model generated 5 outputs for the same prompt, which led to a total of 945 outputs  (27 reference sites, 7 models, 5 repeats).

\subsection{Trace Generation}

We used UI-TARS-1.5-7B~\cite{qin2025ui}, a vision-based CUA to capture UI states in reference and generated websites during task execution.
We chose to use a vision-based agent, as opposed to one with access to the browser DOM, as we hypothesized it would more closely reflect real user perception and interaction with the UI.

At the beginning of each trace capture, a full-screen browser was launched a headless display server.
The browser was pointed to the live website URL, or to a locally hosted build for generated outputs.
The CUA was run using the officially provided library for UI-TARS, which operated the keyboard, mouse, and took screenshots for up to 50 steps.
We used all default parameters except for temperature, which we increased from 0 to 0.7 for generating different traces.

\subsection{Score Computation}
A UI is scored by measuring the average similarity of its traces with the corresponding traces of a reference.
To compute trace similarity, we selected three examples of reference-based scores found in NLP representing align-based, overlap-based, and distribution-based approaches.

\textbf{Featurization.} Each raw trace is converted into a sequence of $D$-dimensional embeddings,
$x = (\mathbf{e}_1,\ldots,\mathbf{e}_T) \in \mathbb{R}^{T \times D}$.
All traces were featurized using an off-the-shelf CLIP-style model finetuned on UIs~\cite{wu2024uiclip} that maps each screenshot $i_t$ to a fixed-length vector
$\mathbf{e}_t \in \mathbb{R}^{D}$.

\textbf{Aggregation. }To account for the variation between the three CUA runs, we computed the metrics for all nine possible pairings and select the best score.

\subsubsection{Alignment-based Scoring}
For our experiments, we chose to use dynamic time warping (DTW) as an example of an alignment-based measurement~\cite{sakoe1978dynamic}.
DTW, an algorithm that first uses dynamic programming to find an optimal alignment through ``warping'' and computing a cost function from the optimal alignment path.

\begin{equation} \label{eq:dtw}
   \operatorname{DTW}(x_{1:n}, y_{1:m})
= \min_{P \in \mathcal{P}} \sum_{(i,j)\in P} c(x_i, y_j) 
\end{equation}
Equation \ref{eq:dtw} describes DTW, where $\mathcal{P}$ refers to all possible ``paths'' between sequences $x$ and $y$, and $c(x, y)$ refers to the cosine distance between two elements $x_i$ and $y_j$.

To prevent unrealistic assignments, we applied Sakoe-Chiba bands of 10\% of the reference trace length.
If no valid warping was found, then the DTW was set to the length of the reference trace.

\subsubsection{Overlap-based Scoring}
Overlap-based metrics compare sequences by the amount of sub-unit overlap (\textit{e.g.,} n-grams).
BLEU~\cite{papineni2002bleu} is a well-known example used for text, and we used eBLEU~\cite{elnokrashy2023ebleu}, a generalization to continuous-valued sequences.
\begin{equation} \label{eq:ebleu}
\operatorname{eBLEU}(x,y)
= b\,\exp\Bigl(\sum_{k=1}^K w_k \log P_k(x,y)\Bigr)
\end{equation}

Equation \ref{eq:ebleu} describes the computation of eBLEU.
$P_k(x,y)$ is the order-k ``soft precision,'' defined as the average, over candidate k-grams in $x$, of the maximum clipped embedding-space similarity to any reference k-gram in $y$.
Following the original paper, we considered k-grams up to length 4 and define $b$ as the length penalty for sequences with over a 50\% difference in length.

\subsubsection{Distribution-based Scoring}
Finally, we used Word Mover's Distance (WMD) as an example of a distribution-based measurement, which measures the the minimal optimal-transport cost to transform one sequence's distribution of features into the other~\cite{kusner2015wmd}.

\begin{equation}~\label{eq:wmd}
\operatorname{WMD}(x_{1:n}, y_{1:m})
= \min_{T \in \mathcal{U}}
  \sum_{i=1}^n \sum_{j=1}^m T_{ij}\,c(x_i, y_j)
\end{equation}
Equation \ref{eq:wmd} describes the computation of WMD given two sequences $x$ and $y$, where $T$ is a transport plan, and $c(x, y)$ is the Euclidean distance between vectors $x$ and $y$.
\section{Feasibility Evaluation}
We collected ratings of generated UIs from human experts and used them to measured alignment of our computed metrics and a MLLM-as-a-judge baseline.
Ratings were collected in an arena-style evaluation~\cite{pmlr-v235-chiang24b}. %

\textbf{Participants.}
For this validation, the experts were two members of the research team who had PhD degrees in human-computer interaction (HCI) and were familiar with evaluation protocols such as usability testing and heuristic evaluation.
The arena evaluation consisted of blind comparisons, reducing the chance of bias towards any specific condition.
Despite the small number of evaluators, we found that our data were sufficient to reach statistical significance in determining the relative performance.

\textbf{Methodology.}
We based our arena interface on existing ones for frontend development~\cite{vichare2025webdev,designarena2025}. %
The arena page displayed a natural language description of a website but not the associated list of tasks, since we wanted raters to test tasks that they thought would be important given the description.
Two websites from different models corresponding to the same randomly sampled prompt were displayed side-by-side in expandable \texttt{<iframe>} elements.
Raters were instructed to interact with each website and test if it correctly implemented common functionality, then make a decision.

In total, we collected 429 unique pairs from human experts to serve as ground truth, which results in roughly 20 comparisons for each possible pairing of the 7 tested generator models (21 pairings).
We computed Elo ratings for each code generator model following existing approaches~\cite{pmlr-v235-chiang24b}.

\textbf{Baseline. }For additional analysis, we implemented an automated judge~\cite{li2025webdevjudge} that prompted an MLLM to choose between a pair of candidates based the modified code and rendered UI screenshot.
To run this judge, we used Claude Opus 4.5, the strongest available model for design tasks at the time of this experiment~\cite{designarena2025}.
We ran this judge on all human-rated pairs and computed Elo scores.

\subsection{Results}

\begin{table}[t]
  \centering
  \scriptsize
  \setlength{\tabcolsep}{3pt}
  \begin{tabular}{lccccc}
    \toprule
    \textbf{Model} & \textbf{Human} & \textbf{DTW} & \textbf{eBLEU} & \textbf{WMD} & \textbf{MLLM} \\
    \midrule
    Claude Sonnet 4.5 & 1342 (1) & 1210 (1) & 1098 (1) & 1146 (1) & 1314 (1) \\
    GPT 5.1 Codex & 928 (5) & 948 (5) & 969 (5) & 940 (5) & 1162 (2) \\
    GPT-OSS 20B & 826 (6) & 909 (7) & 957 (7) & 915 (7) & 852 (7) \\
    GPT-OSS 120B & 776 (7) & 1021 (2) & 999 (3) & 928 (6) & 858 (6) \\
    Gemini Pro 2.5 & 1048 (3) & 1008 (3) & 1024 (2) & 1033 (3) & 874 (5) \\
    Qwen3Coder 30B & 972 (4) & 964 (4) & 998 (4) & 983 (4) & 942 (4) \\
    Qwen3Coder 480B & 1119 (2) & 944 (6) & 967 (6) & 1052 (2) & 996 (3) \\
    \midrule
    Spearman $\rho$ & N/A & 0.25 & 0.39 & \textbf{0.96} & 0.71 \\
    Cohen's $\kappa$ & N/A & 0.37 & 0.23 & \textbf{0.46} & 0.44 \\
    Agreement rate & N/A & 68.1\% & 61.1\% & \textbf{73.0\%} & 71.1\% \\
    \bottomrule
  \end{tabular}
  \caption{Models Elo score (and rank) on the 429 unique pairs rated by human experts. WMD achieves the best Spearman correlation $\rho$ to human rankings.}
  \label{tab:ranking_comparison_with_gt}
\end{table}

Table \ref{tab:ranking_comparison_with_gt} shows the Elo scores computed from human ratings, reference-based metrics, and MLLM judge.
To measure the degree to which a metric aligned with human judgment, we adopted three approaches i) computing the Spearman rank correlation ($\rho$), rater reliability (Cohen's $\kappa$), and iii) the binary agreement rate.

Based on all approaches, WMD was the most effective, with near perfect rank correlation, $\kappa = 0.46$, and 73.0\% agreement rate.
Other reference-based metrics aligned more closely to human labels than the baseline, and while not perfect, may offer other diagnostic utility, e.g., a DTW's alignment path reveals redundant steps in a flow.
The MLLM was the second-best performer in all metrics.
One practical drawback of the MLLM judge was that tested websites often contained thousands of lines of JavaScript code which could present challenging ``needle-in-a-haystack'' problems for larger code-bases and incur significant costs (e.g., hundreds of dollars).
Both FlowEval with the WMD metric and the MLLM judge achieved moderate agreement with human labels~\cite{landis1977measurement}.
This level of agreement is comparable to inter-rater reliability reported among human judges in prior UI evaluation tasks~\cite{wu2024uiclip,duan2024uicrit}, suggesting that the results are reasonable given the subjectivity of the task.

Overall, these results provide evidence that FlowEval, paired with WMD, can provide a more accurate alternative to costlier black-box MLLM judges.
\section{Conclusion}
We introduced FlowEval, a reference-based framework that evaluates generated UIs by comparing their interaction flows against those from reference websites.  
We show that reference-based metrics align strongly with expert judgments and outperform existing MLLM judges.
FlowEval enables a form of automated yet reliable evaluation and can provide more interpretable feedback.

\section*{Limitations}
We discuss the limitations of FlowEval's intended scope and the implementation described in this paper.

\paragraph*{Scope.}
UI quality is multi-dimensional and includes functional correctness, visual aesthetics, accessibility, performance, and overall user experience.
FlowEval focuses primarily on task support, meaning whether a generated UI enables users to complete validated interaction flows.
We view this as an underexplored axis for automated UI evaluation.
As a result, FlowEval should be seen as complementary to prior approaches that emphasize compilation, layout quality, or visual appeal, and future work could combine reference-based flow metrics with these other signals to provide a more holistic assessment.

Reference-based evaluation also has inherent coverage limitations.
High quality designs and experiences require creativity, and creative excellence may require designs that are intentionally novel or unconventional, which can make it difficult to identify an appropriate existing reference.
Currently, we envision FlowEval as an automated approach to evaluating coding agents' ability to implement common UI flows and experiences.

FlowEval evaluates UIs in a fully automated setting, which requires generated websites to be produced without interactive feedback.
In realistic development workflows, interfaces often go through multiple rounds of human iteration and refinement, which can substantially improve quality.
Our results primarily characterize performance under a one-shot generation setting, not the quality achievable with human-in-the-loop iteration, although we feel that one-shot implementation performance is a reasonable proxy for more guided workflows.

\paragraph*{Implementation.}
Our implementation of the FlowEval evaluation framework is aimed to show feasibility of automated, reference-based flow evaluation.
Although not tested, FlowEval could be extended to UIs outside of the web domain and include additional reference-based metrics beyond the three exemplars in our paper.

In our experiments, we sourced the tasks from agent navigation benchmarks, which may be biased toward publicly accessible flows.
Many benchmarks under-represent stateful or restricted workflows, such as those requiring authentication, payments, or other gated interactions.
Benchmark designers could curate a more challenging and representative set of tasks and expand coverage to a broader range of websites, workflows, and domains.

Our evaluation prompts are also more structured than many real-world requests to agents or generative models, which are often underspecified and do not enumerate explicit task requirements. Performance under FlowEval may therefore be viewed as an upper bound on models' ability to infer and support realistic tasks from ambiguous prompts. We hypothesize that performance in this constrained setting is predictive of performance in more naturalistic settings, but validating this relationship is an open problem.

Our experiments were also conducted with the best available APIs and agents available at the time of our experiments.
The coding and navigation performance of LLMs are constantly changing and can affect the metrics computed from our method.
A weak CUA may not be able to successfully execute a task, even if the website's implementation supports it.
To some degree, we expect a CUA to maintain consistent performance across websites to mitigate this effect, e.g., failing the same task on both the reference and generated UI would still result in similar traces.
However, we see an important avenue for future work in studying the robustness of our approach across multiple CUA agents.

\paragraph*{Evaluation.}
Our paper presents an initial feasibility study of our approach, where we recruited two human annotators and generated interaction traces with one CUA model.
Our protocol included measures for reducing the effect of potential annotator bias (blind comparisons) and CUA-specific performance (best-of-n sampling and aggregation). However, our small-scale evaluation may not represent the same performance characteristics as larger evaluator pools and the behavior of stronger CUA models.
In future work, we plan to perform a more rigorous evaluation of our approach before releasing it as a validated benchmark.

\paragraph*{Generative AI Safety.}
Finally, FlowEval relies on generative models, including large language models and code-generating systems, which can produce harmful or unsafe outputs. This includes risks associated with unsafe code generation, policy-violating content, and agent behaviors that could lead to unintended actions during navigation. Practical deployments require careful sandboxing, content filtering, and safety guardrails.

\bibliography{custom}

\appendix

\section{UI Generation Prompt}
\label{sec:appendix_generation_prompt}

We generated React.js websites with several coding agents using a one-shot prompt.
Our one-shot generation prompt was based on pubicly available prompts used for UI generation and benchmarking~\cite{designarena2025,akhaliq_anycoder_2025}.
We modified existing prompts specifically for generating code that could easily be staged in our React.js harness (\textit{e.g.,} only modifying files in a certain directory and using only certain libraries).

\begin{lstlisting}
You are an expert on frontend design, you will always respond to web design tasks.
Your task is to create a website according to the user's request using the React framework.
When implementing the website, you should follow these rules:
[Implementation Rules]
1. You should use React by default.
2. If the user requires a library that is not installed in current react environment, please use HTML and tell the user the reason.
3. Your task is to create a website using React JSX, where the entire implementation is contained within the src/ directory. src/App.jsx is the main view and it should should export a component called App. Besides the new files you create in the src/ directory, you must not change or edit any other files.

## Common Design Principles

### General Design Guidelines:
- Create a stunning, contemporary, and highly functional website based on the user's request
- Implement a cohesive design language throughout the entire website/application
- Choose a carefully selected, harmonious color palette that enhances the overall aesthetic
- Create a clear visual hierarchy with proper typography to improve readability
- Incorporate subtle animations and transitions to add polish and improve user experience
- Ensure proper spacing and alignment using appropriate layout techniques
- Implement responsive design principles to ensure the website looks great on all device sizes
- Use modern UI patterns like cards, gradients, and subtle shadows to add depth and visual interest
- Incorporate whitespace effectively to create a clean, uncluttered design
- For images, use placeholder images from services like https://placehold.co/     

## React Design Guidelines

### Implementation Requirements:
- You MUST export an **`App`** component as default export in src/App.jsx
- DO NOT include any external libraries, frameworks, or dependencies outside of what is already installed
- Utilize TailwindCSS for styling, focusing on creating a visually appealing and responsive layout
- Do not define any new SVGs - you should use icons from FontAwesome instead from the packages: @fortawesome/fontawesome-svg-core @fortawesome/free-solid-svg-icons @fortawesome/react-fontawesome. You can assume that these packages are installed even though you cannot see them in a package.json in the same directory.
- Avoid using arbitrary values (e.g., `h-[600px]`). Stick to Tailwind's predefined classes for consistency
- Use static data instead of making HTTP requests or API calls to external services. Define all data classes and placeholder data structures in a location so that it can be referenced when creating components. Since a lot of the website's functionality may depend on this static data, you should think carefully about what should be inside of it, define it early, and hook it up to various parts of the app.
- Utilize Tailwind's typography classes to create a clear visual hierarchy and improve readability
- Ensure proper spacing and alignment using Tailwind's margin, padding, and flexbox/grid classes
- All functionality, such as buttons, links, search functionality, and forms should be implemented.
- Do not treat this as the starting point for an app - it should be the final complete UI.
- Remember to include alt text for all images.
- There should absolutely NEVER be any dead links or buttons that aren't implemented.

Create a website with the following description: <App Name>: <App Description>. The website must support the following tasks with the entire flows realistically mocked:

- Task 1
- Task 2
- Task n

After implementing those tasks, do the following:
- Imagine what other functionality the website is missing. Be as comprehensive as possible. Think of all the possible reasons why a user might visit this website. Add that to your todo list
- Implement mocked flows for those functionality.
- Make sure all remaining buttons and interactive elements on the website should be mocked up and respond to interactions realisitically.

After you are done, review the code you have generated and ensure it meets all of the requirements. Do not run the server or install additional packages. Never execute the `npm` command.
\end{lstlisting}

\section{Arena Labeling Interface}
\begin{figure*}[!htb]
    \centering
    \includegraphics[width=\linewidth]{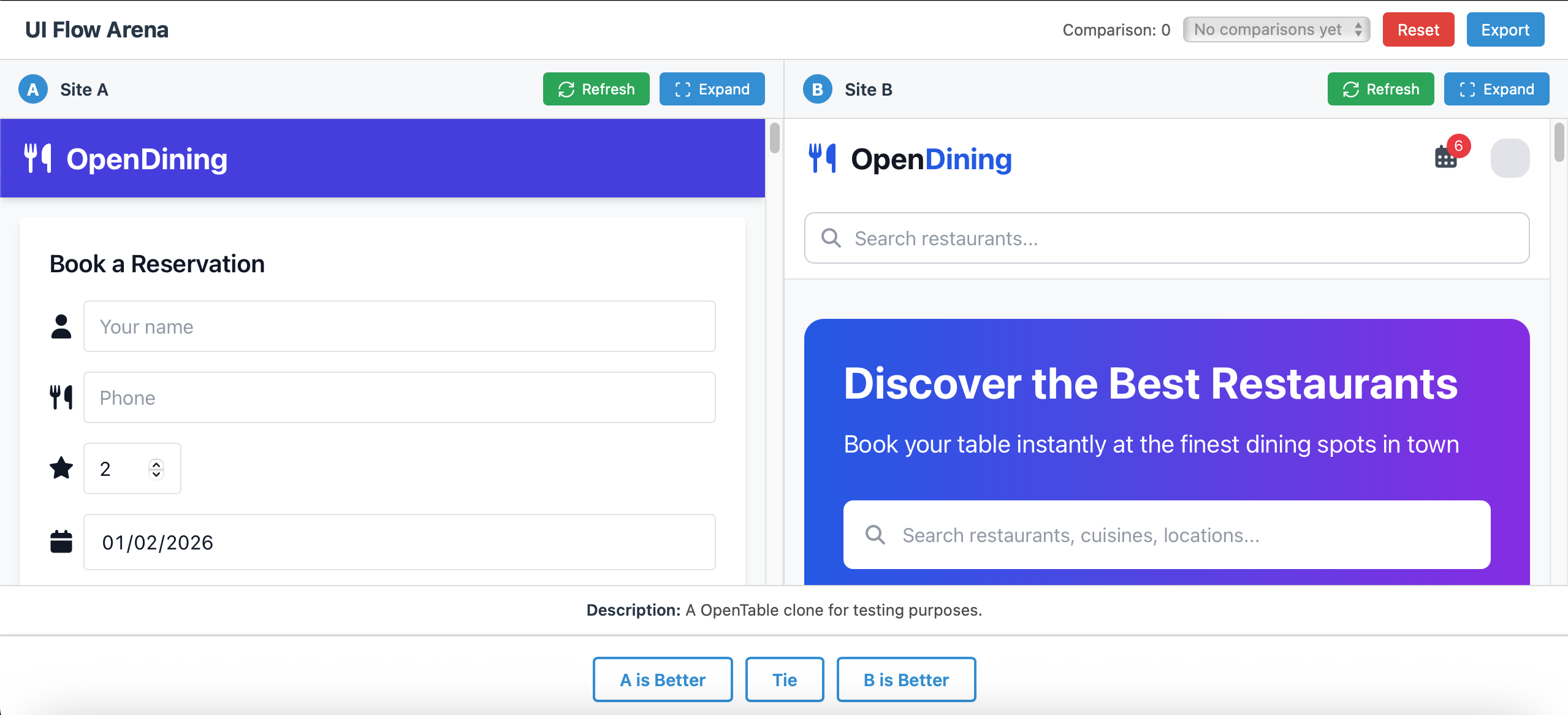}
    \caption{A screenshot of our arena interface used by human raters in our experiment.}
    \label{fig:arenainterface}
\end{figure*}

Figure \ref{fig:arenainterface} shows the rating interface used by human raters during our experiment. It is based on the design of existing interface used for rating UIs~\cite{vichare2025webdev,designarena2025}.

\section{MLLM Judge Prompt}
Following previous work~\cite{li2025webdevjudge}, we employed an MLLM judge in pair mode. For each comparison, the assignment of candidates to Model A and Model B was randomized to mitigate positional bias.

Each request included: i) the user query (app name + description), ii) Code A and Code B (full contents of files that were \textit{new or modified} relative to the boilerplate), and iii) two screenshots labeled ``Initial State A" and ``Initial State B."

The judge prompt followed the publicly released WebDevJudge Likert rubric~\cite{li2025webdevjudge}, with output required as JSON scores for A/B per sub-criterion:

\begin{lstlisting}
{
  "1.1": {"A": 1-5, "B": 1-5},
  ...,
  "4.3": {"A": 1-5, "B": 1-5}
}
\end{lstlisting}

We then converted scores to a pairwise decision by summing criterion scores and choosing the model with the higher score.

\section{Raw Metric Values}
Our main experiment computed the reference-based metrics for each comparison pair and used that to create a binary label (Table \ref{tab:ranking_comparison_with_gt}). These binary labels were used to calculate an Elo score that was comparable to our human ranking data. Table \ref{tab:appendix_metric_scores} uses a difference approach were the raw metric values are aggregated (median) across all model outputs, which is sorted to generate a ranking. We found that this approach did not perform as well as the Elo method, but we include the values here for reference.
\begin{table}[t]
  \centering
  \resizebox{\columnwidth}{!}{%
    \begin{tabular}{lccc}
      \toprule
      \textbf{Model} & \textbf{DTW} $\downarrow$ & \textbf{eBLEU} $\downarrow$ & \textbf{WMD} $\downarrow$ \\
      \midrule
      Claude Sonnet 4.5 & 9.550 (2) & \textbf{0.226 (1)} & \textbf{0.603 (1)} \\
      Qwen3Coder 480B & 15.014 (7) & 0.355 (4) & 0.640 (2) \\
      Gemini Pro 2.5 & \textbf{8.636 (1)} & 0.319 (2) & 0.668 (4) \\
      Qwen3Coder 30B & 14.902 (6) & 0.383 (7) & 0.663 (3) \\
      GPT 5.1 Codex & 13.788 (5) & 0.371 (5) & 0.691 (7) \\
      GPT-OSS 20B & 13.405 (4) & 0.381 (6) & 0.687 (6) \\
      GPT-OSS 120B & 11.542 (3) & 0.345 (3) & 0.686 (5) \\
      \midrule
      Spearman $\rho$ vs GT rank & 0.04 & 0.46 & \textbf{0.82} \\
      \bottomrule
    \end{tabular}%
  }
  \caption{Reference metric medians (rank in parentheses) on the comparison pairs rated in our experiments. Spearman $\rho$ is computed against rankings derived from human Elo score. WMD remains the best performing metric.}
  \label{tab:appendix_metric_scores}
\end{table}

\end{document}